\documentclass[aps,pra,preprint,superscriptaddress,longbibliography]{revtex4-1}

\usepackage{amsmath}
\usepackage{amsthm}
\usepackage{amsfonts}
\usepackage{amssymb}
\usepackage{xcolor,graphicx}
\usepackage{tikz}
\usepackage{color}
\usepackage[pdftex]{hyperref} 
\usepackage{longtable}
\usepackage{array}
\usepackage{dsfont}

\begin{document}
	
	\title{$\mathcal{PT}$-symmetry from Lindblad dynamics in an optomechanical system}	

	\author{B. Jaramillo \'Avila}
	\affiliation{CONACYT-Instituto Nacional de Astrof\'{i}sica, \'{O}ptica y Electr\'{o}nica, Calle Luis Enrique Erro No.~1. Sta. Ma. Tonantzintla, Pue. C.P. 72840, M\'{e}xico.}

	\author{C. Ventura-Vel\'{a}zquez}
	\affiliation{Instituto Nacional de Astrof\'{i}sica, \'{O}ptica y Electr\'{o}nica, Calle Luis Enrique Erro No.~1. Sta. Ma. Tonantzintla, Pue. C.P. 72840, M\'{e}xico.}
	
	\author{R. de J. Le\'on-Montiel}
	\affiliation{Instituto de Ciencias Nucleares, Universidad Nacional Aut\'onoma de M\'exico, Apartado Postal 70-543, 04510 Cd. Mx., M\'{e}xico.}

	\author{Y. N. Joglekar}
	\affiliation{Department of Physics, Indiana University Purdue University Indianapolis (IUPUI), Indianapolis, Indiana 46202 USA.}
	
	\author{B. M. Rodr\'iguez-Lara}
	\affiliation{Tecnologico de Monterrey, Escuela de Ingenier\'ia y Ciencias, Ave. Eugenio Garza Sada~2501, Monterrey, N.L., M\'exico, 64849.}
	\affiliation{Instituto Nacional de Astrof\'{i}sica, \'{O}ptica y Electr\'{o}nica, Calle Luis Enrique Erro No.~1. Sta. Ma. Tonantzintla, Pue. C.P. 72840, M\'{e}xico.}
	
	\date{\today}
	
	\begin{abstract}
	The optomechanical state transfer protocol provides effective, lossy, quantum beam-splitter-like dynamics where the strength of the coupling between the electromagnetic and mechanical modes is controlled by the optical steady-state amplitude. By restricting to a subspace with no losses, we argue that the transition from mode-hybridization in the strong coupling regime to the damped-dynamics in the weak coupling regime, is a signature of the passive parity-time ($\mathcal{PT}$) symmetry breaking transition in the underlying non-Hermitian quantum dimer. We compare the dynamics generated by the quantum open system (Langevin or Lindblad) approach to that of the $\mathcal{PT}$-symmetric Hamiltonian, to characterize the cases where the two are identical. Additionally, we numerically explore the evolution of separable and correlated number states at zero temperature as well as thermal initial state evolution at room temperature. Our results provide a pathway for realizing non-Hermitian Hamiltonians in optomechanical systems at a quantum level.
	\end{abstract}
	
	
	\maketitle

\section{Introduction}
Photonics provides a fertile ground for the classical simulation of non-Hermitian systems with gain, loss, or both, including systems with balanced gain and loss, i.e. parity-time ($\mathcal{PT}$) symmetric systems \cite{ElGanainy2018}. In such a simulation with classical light, the complex potentials in the $\mathcal{PT}$-symmetric Hamiltonian of a quantum system translate into complex refractive media that represent localized amplification or absorption. These parity-time symmetric structures are described by a Schr\"odinger-like differential equation, where the renormalized paraxial propagation mimics quantum dynamics of a non-relativistic particle in the presence of complex optical potentials~\cite{Ruschhaupt2005,ElGanainy2007,Huerta2016}. A key feature of the $\mathcal{PT}$-symmetric Hamiltonian is that at small gain-loss strength, its spectrum remains purely real, its linearly independent eigenfunctions are no longer orthogonal, but continue to remain simultaneous eigenfunctions of the combined $\mathcal{PT}$ operator. When the gain-loss strength is large, the spectrum renders into complex conjugate eigenvalue pairs, and the associated eigenfunctions transform into the other under the $\mathcal{PT}$ operation \cite{Joglekar2013}. This transition from the $\mathcal{PT}$-symmetric phase to the $\mathcal{PT}$-symmetry broken phase occurs at an exceptional point (EP) where the algebraic multiplicity of the Hamiltonian differs from its geometric multiplicity \cite{Kato1995}. The dynamics of non-Hermitian systems across and in the neighborhood of the transition point have been extensively investigated in recent years in mostly classical, optical realizations.

On a fundamental level, the effective, non-Hermitian Hamiltonian model ignores the thermal fluctuations attendant with the loss (due to fluctuation-dissipation theorem \cite{Kubo1966}) and zero-temperature quantum fluctuations attendant with the gain (due to the vacuum noise in linear quantum amplifiers \cite{Caves1982}). Therefore, non-Hermitian dynamics has been realized in mode-selective lossy systems, including heralded single photons \cite{Xiao2017}, ultracold atoms \cite{Li2019}, and superconducting transmons \cite{Naghiloo2019}, where the thermal fluctuations can be safely ignored.
Such lossy systems are also a promising candidate for observing $\mathcal{PT}$-symmetric quantum optics across EPs of arbitrary order with appropriate post-selection~\cite{Quiroz2019}. In systems with both gain and loss, the inclusion of non-classical light requires the introduction of (quantum) fluctuations induced by the linear media either by Langevin equation~\cite{Agarwal2012,Huerta2017,Longhi2018} or Lindblad master equation~\cite{Scheel2018,Schomerus2010,Perinova2019} formalism. Indeed, the trace-preserving, steady-state generating Lindblad approach allows us to understand, in a more realistic way, the dynamics of optomechanical systems and, at the same time, the emergence of a non-Hermitian Hamiltonian in this approach.

In this paper, we provide a thorough analysis of both approaches and present the main differences between them. As model system, we consider a standard, first red-sideband, strongly-driven optomechanical system, where the optomechanical coupling leads to the hybridization of the electromagnetic and mechanical modes~\cite{Dobrindt2008}. This protocol generates an effective, linearized quantum-fluctuation Hamiltonian for the electromagnetic and mechanical modes that is equivalent to that of a lossy, quantum beam-splitter for the two modes~\cite{Aspelmeyer2014}.

The plan for the paper is as follows. First, we introduce the basic model and its Lindblad dynamics, recall the corresponding Langevin equation treatment, and obtain the mode-selective lossy Hamiltonian. We show that coupling to a thermal reservoir leads to a passive $\mathcal{PT}$-symmetric dimer dynamics where the electromagnetic driving controls the $\mathcal{PT}$-symmetric or $\mathcal{PT}$-symmetric broken phases of the dimer. Next, we present numerical results that compare the non-Hermitian evolution of the density matrix with the evolution under a zero-temperature Lindblad master equation for product initial states and correlated $N00N$ initial states. Then, we present finite temperature results for the transition from strong to weak coupling regimes in state transfer protocol at finite temperature to relate it with the $\mathcal{PT}$-symmetry transition. We conclude the paper with a brief discussion.

\section{Results}
\subsection{Optomechanical state-transfer protocol}

The Hamiltonian for the standard optomechanical system \cite{Pace1993,Law1995}, in a frame rotating at the pump frequency $\omega_p$ and units of $\hbar$,
\begin{eqnarray}
	\label{Eq:H0}
	\hat{H}_{0} &=& \left(\omega_{a}-\omega_{p}\right) \hat{a}^{\dagger}\hat{a} +\omega_{b}  \hat{b}^{\dagger}\hat{b}
	+  g_{0}  \hat{a}^{\dagger}\hat{a} (\hat{b}^{\dagger}+\hat{b}) +\Omega(\hat{a}^{\dagger} + \hat{a})/2,
\end{eqnarray}
models the interaction of an electromagnetic mode, with frequency $\omega_a$ and annihilation operator $\hat{a}$, and a mechanical mode, with frequency $\omega_b$ and annihilation operator $\hat{b}$.
The bare optomechanical coupling, $g_0$, indicates the coupling between the dimensionless intensity of the electromagnetic mode, provided by the number operator $\hat{a}^{\dagger} \hat{a}$, and the dimensionless mechanical displacement, $(\hat{b}^\dagger+\hat{b})$.
The parameter $\Omega$ gives the strength of the electromagnetic pump.
Hereafter, we will use subscripts $a$ and $b$ to label electromagnetic and mechanical modes, respectively. Strong driving allows us to split the mode dynamics into semi-classical and quantum fluctuation parts, $\hat{a}=\alpha +\hat{c} $ and $\hat{b}= \beta + \hat{d}$~\cite{Mancini1994,Paternostro2006}. In the presence of a thermal bath, which introduces dissipation for both modes, the semi-classical part shows a steady state with electromagnetic coherent amplitude $\alpha = -i\Omega / 2\left[(\omega_{a}-\omega_{p}) -i\gamma_a/2\right]$ and mechanical coherent amplitude $\beta= - g_{0} \vert \alpha \vert^{2}/\left[\omega_b-i\gamma_{b}/2\right]$.
Here $\gamma_{a}$ and $\gamma_{b}$ are the phenomenological decay rates for the electromagnetic and mechanical mode-occupation numbers, respectively.
Under red-sideband driving, $\omega_p=\omega_a - \omega_b + 2g_0\Re(\beta)$, and the rotating-wave approximation, a quantum beam-splitter Hamiltonian provides the dynamics for the quantum fluctuation,
\begin{eqnarray}
	\label{Eq:BeamSplitterHamiltonian}
	\hat{H} &=& \omega_{b} (\hat{c}^{\dagger}\hat{c} + \hat{d}^{\dagger}\hat{d})+g(\hat{c}^{\dagger}  \hat{d} + \hat{c} \hat{d}^{\dagger}),
\end{eqnarray}
where the steady-state electromagnetic coherent amplitude enhances the bare optomechanical coupling, $g=g_0|\alpha|$ \cite{Genes2008}.

In this scenario, the Lindblad master equation \cite{Breuer2007,Wiseman2009},
\begin{eqnarray}
	\label{Eq:Lindblad}
	\partial_t\hat{\rho} &=& i[\hat{\rho},\hat{H}]+\gamma_a\bar{n}_a\mathcal{D}[\hat{c}^\dagger]\hat{\rho}+\gamma_a(\bar{n}_a+1)\mathcal{D}[\hat{c}]\hat{\rho}
	+ \gamma_b\bar{n}_b\mathcal{D}[\hat{d}^\dagger]\hat{\rho}+\gamma_b(\bar{n}_b+1)\mathcal{D}[\hat{d}]\hat{\rho},
\end{eqnarray}
governs the dynamics of the optomechanical density matrix, $\hat{\rho} \equiv \hat{\rho}_{ab}(t)$, coupled to a thermal bath defined by the action of the zero-trace superoperator
\begin{eqnarray}
	\label{Eq:superoperator}
	\mathcal{D}[\hat{A}]\hat{\rho} &=& \hat{A}\hat{\rho}\hat{A}^{\dagger}-\frac{1}{2}\left[\hat{A}^{\dagger}\hat{A}\hat{\rho}+\hat{\rho}\hat{A}^{\dagger}\hat{A}\right],
\end{eqnarray}
where the average thermal mode-occupation numbers,  $\bar{n}_{x}=1/(e^{\omega_{x}/k_BT}-1)$ with $x=\{a,b\}$, are given in terms of Boltzmann constant $k_B$ and the bath temperature $T$.
The anti-commutator term in Eq.(\ref{Eq:superoperator}) can be interpreted as a purely imaginary gain or loss potential in an effective, non-Hermitian Hamiltonian.
At zero temperature, the Lindblad approach leads to the following equations for the average excitation numbers $\langle \hat{c}^\dagger \hat{c} \rangle$ and $\langle \hat{d}^\dagger \hat{d}\rangle$,
\begin{eqnarray}
	\label{Eq:Lind1}
	\partial_{t} \langle \hat{c}^{\dagger} \hat{c}  \rangle &=& +2 g \textrm{ Im}\langle \hat{c}^{\dagger} \hat{d} \rangle -\gamma_{a} \langle \hat{c}^{\dagger} \hat{c}  \rangle, \\
	\label{Eq:Lind2}
	\partial_{t} \langle \hat{d}^{\dagger} \hat{d}  \rangle &=& -2 g \textrm{ Im} \langle \hat{c}^{\dagger} \hat{d} \rangle-\gamma_{b} \langle \hat{d}^{\dagger} \hat{d} \rangle.
\end{eqnarray}

The quantum Langevin equations of motion for the annihilation operators~\cite{Ventura2019},
\begin{eqnarray} \label{Eq:LangevinFull}
	i\partial_{t} \left[\begin{array}{c} \hat{c}\\\hat{d} \end{array}\right]
	&=&
	\left[\begin{array}{cc} \omega_{b} -i\gamma_{a}/2	& g \\ g & \omega_{b} -i\gamma_{b}/2 \end{array} \right] \left[\begin{array}{c} \hat{c}\\ \hat{d} \end{array}\right] -i\left[\begin{array}{c}\hat{\xi}_{a} \\ \hat{\xi}_{b} \end{array}\right],
\end{eqnarray}
provide an equivalent approach to the open quantum evolution. Here, the dimensionful operators $\hat{\xi}_{x}$ with zero mean and correlation functions $\langle \hat{\xi}_{x}^{\dagger}(t) \hat{\xi}_{x}(s) \rangle = \gamma_x \bar{n}_{x} \delta(t-s)$ and $\langle \hat{\xi}_{x}(t) \hat{\xi}_{x}^{\dagger}(s) \rangle = \gamma_x (\bar{n}_{x}+1) \delta(t-s)$, with $x=\{a,b\}$, model the quantum noise for the electromagnetic and mechanical modes respectively. A Hamiltonian with specific mode losses,
\begin{eqnarray}
	\label{Eq:HL}
	\hat{H}_{L} &=& \hat{H}-i( \gamma_{a} \hat{c}^{\dagger}\hat{c} + \gamma_{b} \hat{d}^{\dagger}\hat{d})/2,
\end{eqnarray}
generates the first term on the right-hand side of Eq.(\ref{Eq:LangevinFull}). When confined to a subspace with a fixed total excitation number $\hat{N}=\hat{c}^\dagger\hat{c}+\hat{d}^\dagger\hat{d}$, Eq.(\ref{Eq:HL}) becomes
$\hat{H}_{L} = \left( \omega_b - i \left[ \gamma_{a} + \gamma_{b} \right]/4 \right) \hat{N}
+( \hat{c}^\dagger\,\,\hat{d}^\dagger)H_{\mathcal{PT}}(\hat{c}\,\,\hat{d})^T$ with
\begin{eqnarray}
	H_{\mathcal{PT}} = g \sigma_{x} - i \Gamma \sigma_{z},
\end{eqnarray}
where $\sigma_{x}, \sigma_{z}$ are standard Pauli matrices and $\Gamma=(\gamma_a-\gamma_b)/4$. It follows that the decay rates of the two eigenmodes of $\hat{H}_L$ are equal ($\mathcal{PT}$-symmetric phase) for $|\Gamma| < g$, they reach the maximum at $|\Gamma|=g$, and a slowly decaying eigenmode emerges for $|\Gamma|>g$~\cite{Quiroz2019,Leon2018,RodriguezLara2015,Li2019,Joglekar2018}.

The dynamics generated by Eq.(\ref{Eq:Lindblad}) and Eq.(\ref{Eq:LangevinFull}) are completely equivalent \cite{Ventura2019}. However, we want to identify and elucidate the cases where they are equivalent to the non-unitary time evolution generated by the non-Hermitian Hamiltonian, i.e. Eq.(\ref{Eq:HL}). In the absence of the quantum noise terms, the non-Hermitian approach gives the following equations of motion for the mode occupation numbers,
\begin{eqnarray}
	\label{Eq:Loss1}
	\partial_{t} \langle \hat{c}^{\dagger} \hat{c}\rangle_L &=& +2g\textrm{ Im}\langle\hat{c}^{\dagger} \hat{d} \rangle_{L}-\langle \hat{c}^{\dagger} \hat{c} (\gamma_a\hat{c}^{\dagger} \hat{c}+\gamma_b\hat{d}^{\dagger} \hat{d})\rangle_{L},\\
	\label{Eq:Loss2}
	\partial_{t} \langle \hat{d}^{\dagger} \hat{d}  \rangle_{L} &=& -2 g\textrm{ Im}\langle \hat{c}^{\dagger} \hat{d} \rangle_{L} -\langle \hat{d}^{\dagger} \hat{d} (\gamma_a\hat{c}^{\dagger} \hat{c}+\gamma_b\hat{d}^{\dagger} \hat{d})\rangle_{L}.
\end{eqnarray}
In the following, we compare the numerical results obtained by solving Eqs.(\ref{Eq:Lind1})-(\ref{Eq:Lind2}) with those from Eqs.(\ref{Eq:Loss1})-(\ref{Eq:Loss2}).
We explore both zero and finite temperatures with initial states that are either product states or correlated $N00N$ states.

\subsection{Numerical results}

For our simulations, we make use of optomechanical parameters from an experimental state transfer protocol  $\{ \omega_{a}, \omega_{b}, \gamma_{a}, \gamma_{b} \} = \{ 1.02 \times 10^{10}, 1.59 \times 10^{7}, 3.26 \times 10^{5}, 3.00 \times 10^{2}  \} $ Hz \cite{Cohen2015}.
The experimental enhanced optomechanical coupling $g = 1.33 \times 10^{-2} ~ \omega_{b} $ provides dynamics in the $\mathcal{PT}$-symmetric regime.
We calculate the required value to reach the exceptional point, $g = (\gamma_{a} - \gamma_{b})/4 = 5.12 \times 10^{-3} ~ \omega_{b} = 8.14 \times 10^{4}$~Hz.
For the broken symmetry regime, we take an order of magnitude less than the reported experimental value, $ g = 1.33 \times 10^{-3} ~\omega_{b} $ without further consideration regarding the validity of the mean-field approximation. For the sake of simplicity, we start our numerical experiments for Lindblad master equation carried at zero temperature and the initial states are given in terms of Fock states. It is important to remark that, even though zero-temperature conditions are ideal for optomechanical experiments, simulations assuming such condition can help elucidate the difference in the dynamics of both approaches, namely the full quantum analysis and the non-Hermitian Hamiltonian approach. For simulations at zero temperature, we use bosonic subspaces of dimension equal to the maximum number of excitations plus two to unfold and reduce the complex differential equations into a set of real differential equations solved using standard Livermore Solver for Ordinary Differential Equations (LSODA) methods. At finite temperature, the solutions to the Langevin equations are obtained exactly by means of an adaptive integrator~\cite{Ventura2019}.

\begin{figure*}[tpb!]
	\centering
	\includegraphics{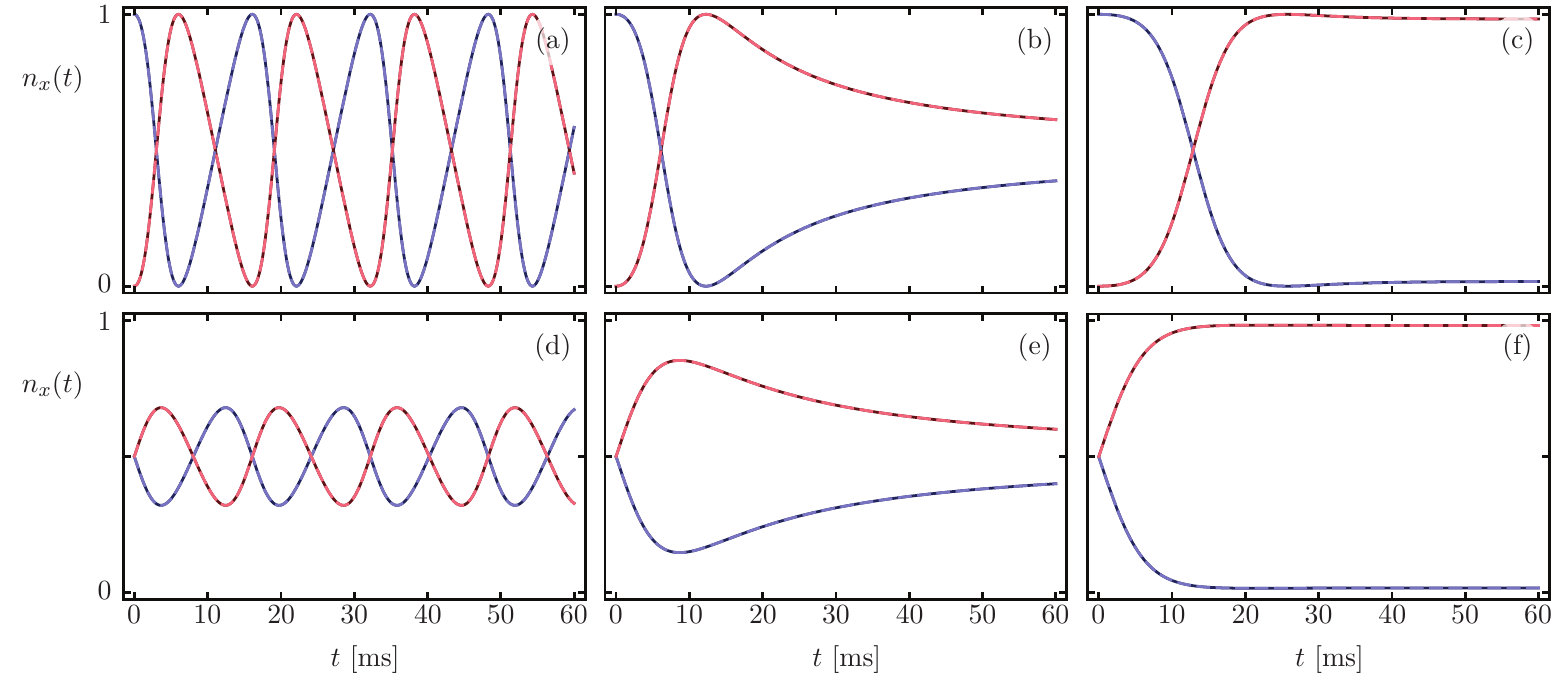}
	\caption{Time-dependent occupation numbers $n_a(t)$ (blue) and $n_b(t)$ (red) for an initial (a-c) separable, $\vert \psi(0) \rangle = \vert 1, 0 \rangle$,  and (d-f) correlated, $\vert \psi(0) \rangle = \left( \vert 1, 0 \rangle + \vert 0,1 \rangle \right)/\sqrt{2}$, single-excitation state. Solid and dashed lines correspond to Lindblad equation and non-Hermitian evolution, in that order.                                  Columns show dynamics in the $\mathcal{PT}$-symmetric region, at the exceptional point, and in the $\mathcal{PT}$-symmetry broken region from left to right.}\label{fig:Fig1}
	\vspace{-3mm}
\end{figure*}

We start with the single-excitation subspace. In this limit, the Lindblad dynamics and the non-Hermitian dynamics are identical,
\begin{eqnarray}
	\partial_{t} \langle \hat{c}^{\dagger} \hat{c}  \rangle &=\;\; \partial_{t} \langle \hat{c}^{\dagger} \hat{c}  \rangle_{L} \;\;=& +2 g \textrm{Im} \langle \hat{c}^{\dagger} \hat{d} \rangle  - \gamma_{a} \langle \hat{c}^{\dagger} \hat{c}  \rangle, \nonumber \\
	\partial_{t} \langle \hat{d}^{\dagger} \hat{d}  \rangle &=\;\; \partial_{t} \langle \hat{d}^{\dagger} \hat{d}  \rangle_{L} \;\;=& -2 g \textrm{Im} \langle \hat{c}^{\dagger} \hat{d} \rangle - \gamma_{b} \langle \hat{d}^{\dagger} \hat{d} \rangle,
\end{eqnarray}
because $\langle \hat{x}^{\dagger} \hat{x} \hat{x}^{\dagger} \hat{x} \rangle = \langle \hat{x}^{\dagger} \hat{x}  \rangle$ for $x=\{a,b\}$, and $\langle \hat{c}^{\dagger} \hat{c} \hat{d}^{\dagger} \hat{d} \rangle = 0$ in the single-excitation subspace. Figure~\ref{fig:Fig1} shows the occupation numbers for the electromagnetic mode $n_a(t)$ and the mechanical mode $n_b(t)$ obtained via the Lindblad master equation (solid lines), and the non-Hermitian evolution (dashed lines). The initial state is separable (first row), and a correlated $N00N$ state (second row). The first, second, and third columns correspond to the system in the $\mathcal{PT}$-symmetric region ($g=1.33\times 10^{-2}\omega_b$), at the exceptional point ($g=5.12\times 10^{-3}\omega_b$), and in the $\mathcal{PT}$-symmetry broken region ($g=1.33\times 10^{-3}\omega_0$) respectively.

To explore the dynamics beyond the single-excitation subspace, we define instantaneously renormalized excitation numbers and first order correlation or coherence,

\begin{eqnarray}
	\label{eq:na}
	n_{a}(t)  & =& \langle \hat{c}^{\dagger}\hat{c}\rangle/\langle \hat{N} \rangle,\\
	\label{eq:nb}
	n_{b}(t)  &= & \langle \hat{d}^{\dagger} \hat{d}\rangle/\langle \hat{N}\rangle,\\
	\label{eq:g1}
	g^{(1)}(t) &=& \langle \hat{c}^{\dagger} \hat{d} \rangle/\langle \hat{N} \rangle.
\end{eqnarray}
We note that the process of instantaneous renormalization is equivalent to restricting to a fixed excitation number $\langle\hat{N}\rangle$ sector. In this sector, the Hamiltonian $\left( \hat{c}^{\dagger}~\hat{d}^{\dagger} \right) H_\mathcal{PT} \left( \hat{c}~\hat{d} \right)^{T}$
is an $N+1$ matrix in the photon-phonon number basis and post-selecting to this sector is equivalent to measuring the quantities $n_a(t), n_b(t)$ and $g^{(1)}(t)$~\cite{Quiroz2019,Naghiloo2019}.

\begin{figure*}[tpb!]
	\centering
	\includegraphics{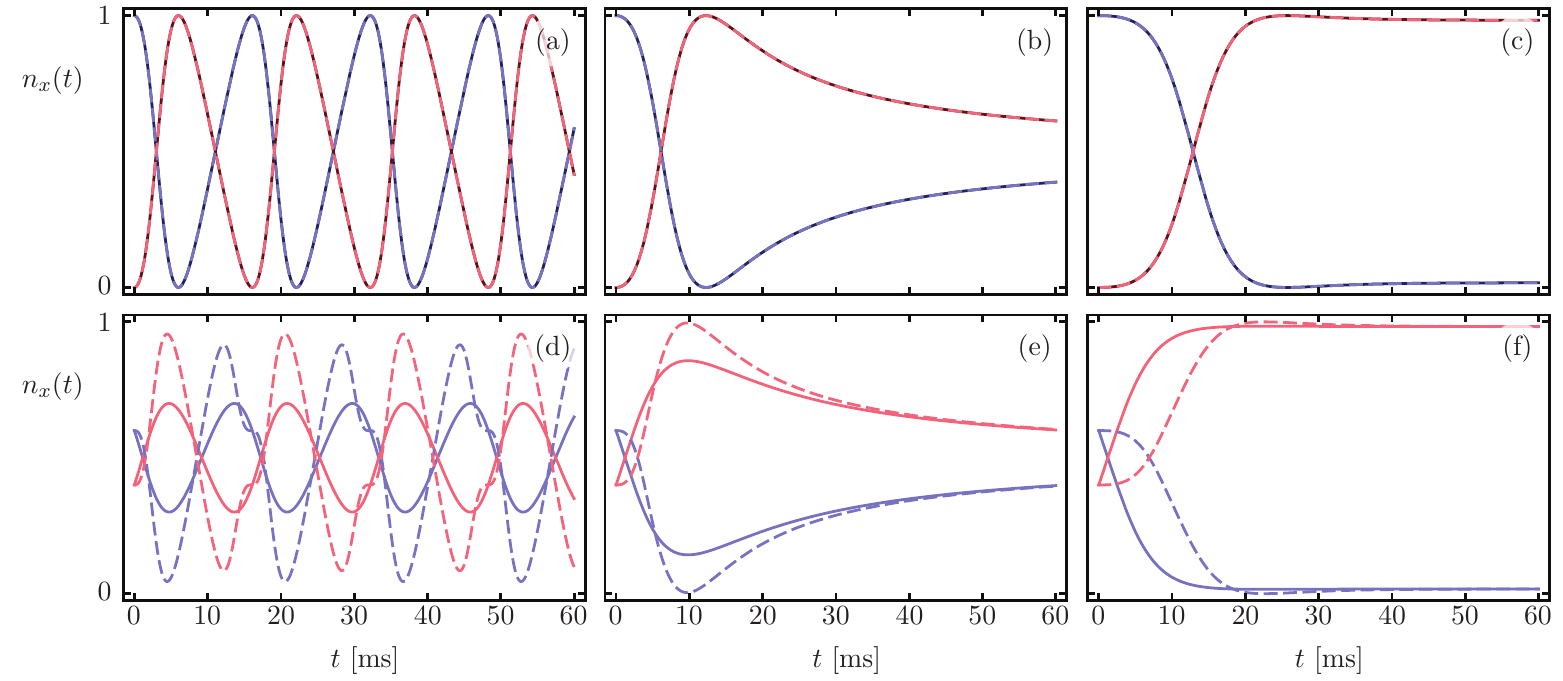}
	\caption{Time dependent occupation numbers $n_a(t)$ (blue) and $n_b(t)$ (red) for initial separable states (a)-(c) $|\psi(0)\rangle=\vert N, 0 \rangle$ with $N=5$ and (d-f) $|\psi(0)\rangle=\vert N-m, m \rangle$ with $N=5$ and $m=2$. Columns show dynamics in the $\mathcal{PT}$-symmetric region, at the exceptional point, and in the $\mathcal{PT}$-symmetry broken region from left to right. Solid and dashed lines correspond to Lindblad equation and non-Hermitian evolution, in that order. The full Lindblad result differs from the non-Hermitian Hamiltonian evolution, but has clear signatures of the $\mathcal{PT}$-symmetry breaking transition.}\label{fig:Fig2}
	\vspace{-3mm}
\end{figure*}

The top row in Fig.~\ref{fig:Fig2} shows occupation numbers $n_a(t), n_b(t)$ for an initial state $|\psi(0)\rangle=\vert N, 0 \rangle$ with $N=5$ obtained from the Lindblad (solid lines) and non-Hermitian (dashed lines) dynamics. The bottom row, on the other hand, shows results for $|\psi(0)\rangle=\vert N-m, m \rangle$ with $m=2$ and $N=5$. We observe the three well defined dynamical regimes: the anharmonic oscillations in $n_x(t)$ have a slightly different period in the $\mathcal{PT}$-symmetric region, but converge asymptotically at the exceptional point and in $\mathcal{PT}$-symmetry broken region. This surprising result, where Lindblad dynamics does not rise to a steady-state behavior, is solely due to the post-selection scheme we have discussed.

\begin{figure*}[tpb!]
	\centering
	\includegraphics{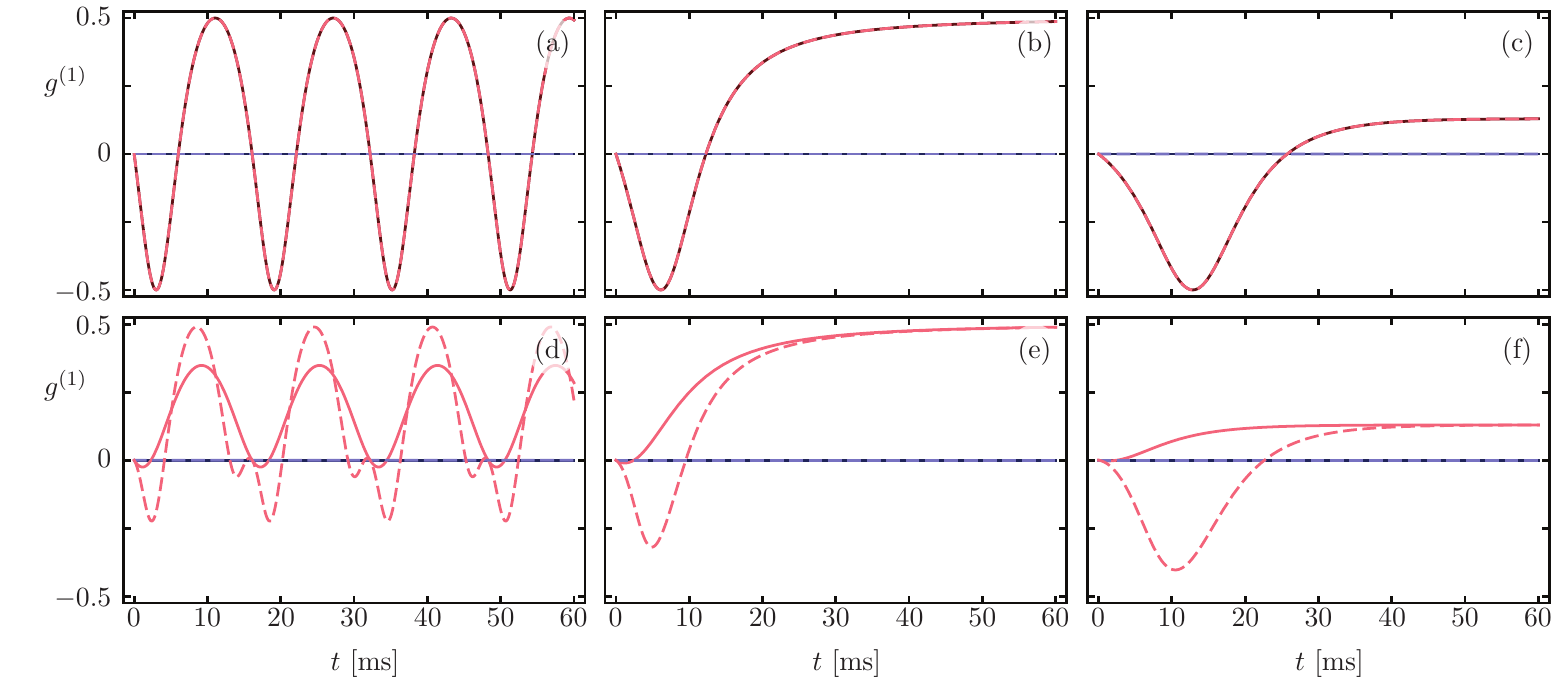}
	\caption{Time-dependent coherence $g^{(1)}(t)$ for initial separable states (a)-(c) $|\psi(0)\rangle=\vert N, 0 \rangle$ and (d-f) $|\psi(0)\rangle=\vert N-m, m \rangle$ with $N=5$ and $m=2$. Blue and red show the real and imaginary part of the optomechanical coherence $g^{(1)}(t)$ respectively. Solid and dashed lines correspond to Lindblad master equation and non-Hermitian evolution, in that order. Columns show dynamics in the $\mathcal{PT}$-symmetric region, at the exceptional point, and in the $\mathcal{PT}$-symmetry broken region from left to right. }\label{fig:Fig3}
	\vspace{-3mm}
\end{figure*}

Figure \ref{fig:Fig3} shows the real (blue) and imaginary (blue) parts of the optomechanical coherence $g^{(1)}(t)$ for separable states $\vert N,0\rangle$ (top row) and $\vert N-m,m\rangle$ (bottom row) respectively. The difference between the Lindblad dynamics (solid lines) and the non-Hermitian Hamiltonian evolution (dashed lines) is again manifest only for product states where both modes are excited.

\begin{figure*}[tpb!]
	\centering
	\includegraphics{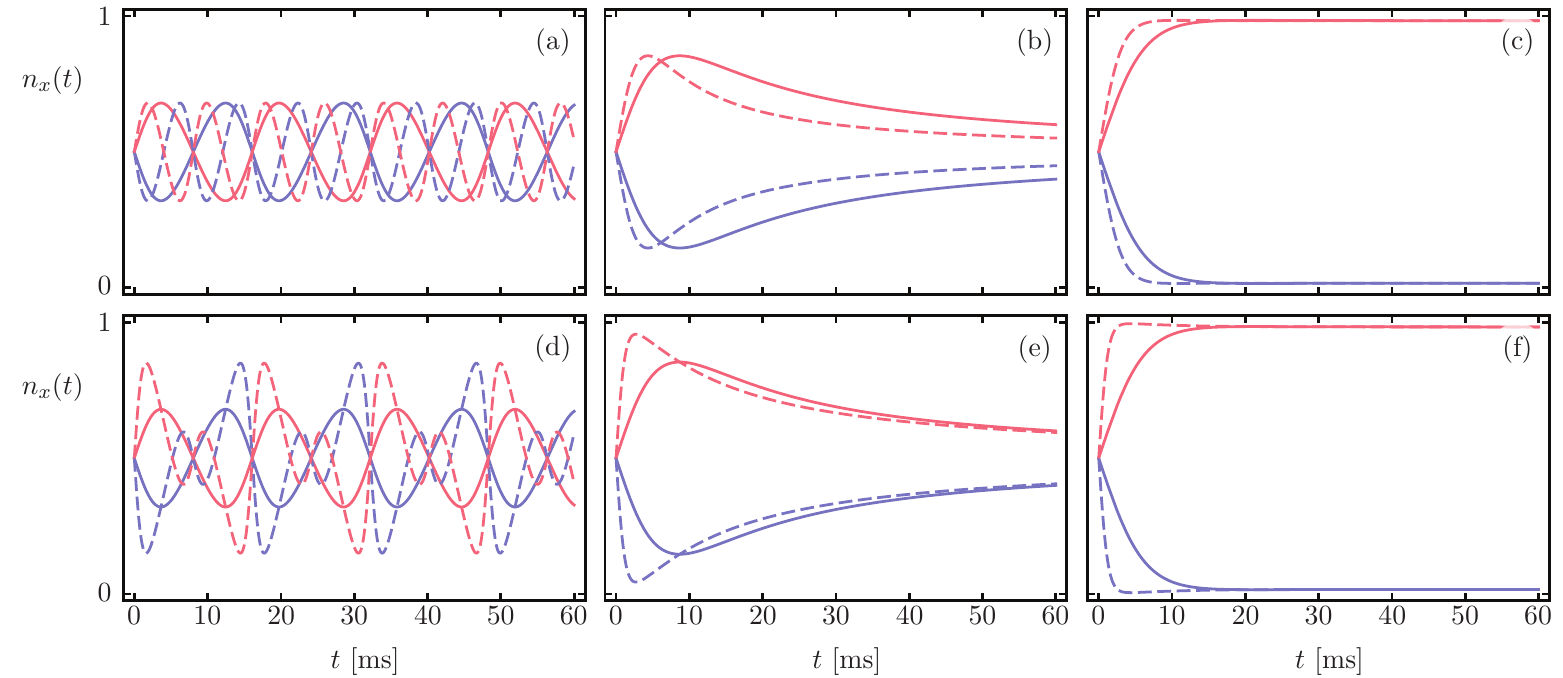}
	\caption{Time dependent occupation numbers $n_a(t)$ (blue) and $n_b(t)$ (red) for correlated $N00N$ state, $|\psi(0)\rangle=\left( \vert N, 0 \rangle + \vert 0,N \rangle \right)/\sqrt{2}$, with (a-c) $N=2$ and (d-f) $N=5$. Solid and dashed lines correspond to Lindblad equation and non-Hermitian evolution, in that order. The full Lindblad result differs from the non-Hermitian Hamiltonian evolution, but has clear signatures of the $\mathcal{PT}$-symmetry breaking transition.}\label{fig:Fig4}
\end{figure*}

Next, we consider the zero-temperature evolution with highly correlated initial states, such as the so-called $N00N$ states, with different values of $N$. Figure~\ref{fig:Fig4} shows that the Lindblad master equation results (solid lines) for the scaled occupation numbers $n_a(t)$ and $n_b(t)$ are independent of $N$, while the non-Hermitian evolution results (dashed line) show deviations that increase with $N$. Again, the characteristic dynamics for the $\mathcal{PT}$-symmetric phase, exceptional point, and $\mathcal{PT}$-symmetry broken phase appear. The anharmonic oscillation period is the same for both approaches in the $\mathcal{PT}$-symmetric region, but the interference in the non-Hermitian evolution differentiates them apart. In the exceptional and broken regimes both dynamics converge asymptotically.

\begin{figure*}[tpb!]
	\centering
	\includegraphics{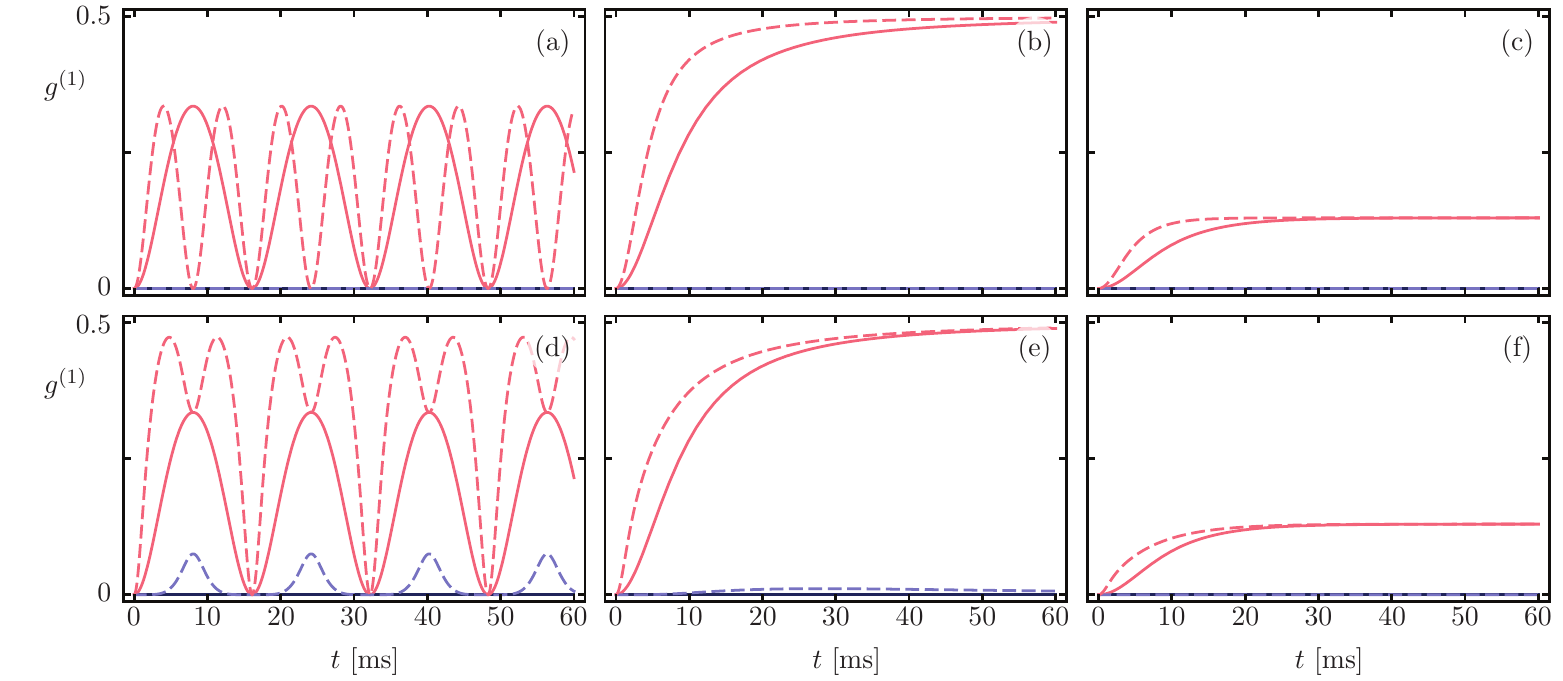}
	\caption{Time-dependent coherence $g^{(1)}(t)$ for correlated $N00N$ state, $|\psi(0)\rangle=\left( \vert N, 0 \rangle + \vert 0,N \rangle \right)/\sqrt{2}$, with (a-c) $N=2$ and (d-f) $N=5$.
		Blue and red show the real and imaginary part of the optomechanical coherence $g^{(1)}(t)$ respectively. Solid and dashed lines correspond to Lindblad equation and non-Hermitian evolution, in that order. Surprisingly, the coherence is maximum at the exceptional point.}\label{fig:Fig5}
\end{figure*}

Figure~\ref{fig:Fig5} shows qualitatively similar results for the optomechanical coherence $g^{(1)}(t)$ with $N00N$ initial states. We find it remarkable that asymptotic value of $g^{(1)}$ at the exceptional point is a maximum in any each case. This is a fascinating effect that could prove useful for preserving coherence in the implementation of quantum information protocols.

\begin{figure*}[htpb!]
	\centering
	\includegraphics{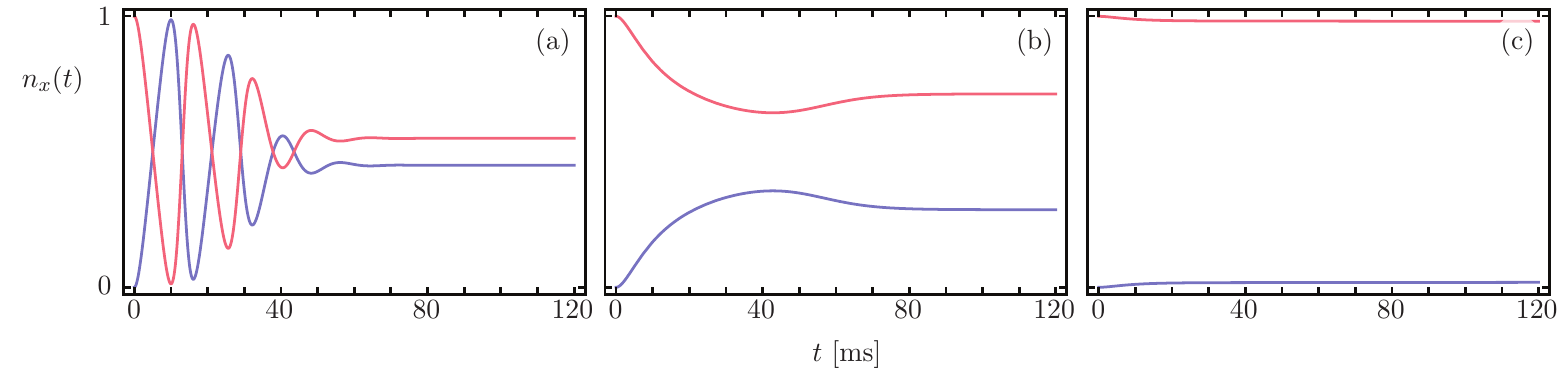}
	\caption{Time dependent occupation numbers $n_a(t)$ (blue) and $n_b(t)$ (red) obtained by solving the Langevin equation at room temperature for initial thermal states. The dynamics has clear signatures of the $\mathcal{PT}$-symmetry breaking transition. }\label{fig:Fig6}
\end{figure*}

Finally, we consider the finite-temperature case that is most relevant to current optomechanical experiments, where the states of the modes are thermal coherent states. In this case, the full quantum dynamics asymptotically provides a thermal steady-state, and the interplay between decay ratios and the enhanced optomechanical coupling provides the dynamics before stabilization~\cite{Dobrindt2008}.
Figure \ref{fig:Fig6} shows these dynamics for finite temperature $T=293$~K where the initial state of the fluctuations given by thermal states with mean excitation numbers $\langle \hat{c}^{\dagger} \hat{c} \rangle = 3.76 \times 10^3$ and $\langle \hat{d}^{\dagger} \hat{d} \rangle = 2.41 \times 10^6$.
For a strong optomechanical coupling, $g > \vert \Gamma \vert$, the electromagnetic and mechanical modes hybridize and this standard mode-splitting results in oscillatory behavior that provides state transfer, Fig.~\ref{fig:Fig6}(a), similar to dynamics in the $\mathcal{PT}$-symmetry region. The transition point from strong to weak coupling occurs at what in non-Hermitian systems is the exceptional point $g = \vert \Gamma \vert$ where power-law approach to steady-state arises and there is no state transfer anymore, Fig.~\ref{fig:Fig6}(b). For weak coupling, $g < \vert \Gamma \vert$, the electromagnetic mode decays according to its damping rate but the mechanical mode shows an effective decay rate that includes the effect of the electromagnetic mode on the mechanical oscillator equivalent to the broken symmetry regime, Fig.~\ref{fig:Fig6}(c). These results are obtained via the full Langevin equation for the same experimental system\cite{Cohen2015}, but now at room temperature.

\section{Conclusion}

We revisited the optomechanical state transfer protocol from a non-Hermitian-Hamiltonian point of view.
After the mean-field approximation, the linearized quantum fluctuation beam-splitter-like Hamiltonian provides us with a theoretical testing ground to compare the results from Lindblad equation and non-Hermitian evolution for a realization of the standard quantum $\mathcal{PT}$-symmetric dimer.

We have shown that Lindblad dynamics  and the non-Hermitian evolution at zero temperature provide identical dynamics for separable initial states where one of the modes is a number state and the other is the vacuum. However, for Fock initial states with non-zero mode numbers, the dynamics are not identical, but continue to be qualitatively similar. The same trend holds for correlated $N00N$ states. Although the zero-temperature bath and Fock initial states cannot be explored in the present-day experimental optomechanical setting, they point to the fact that these regimes are differentiable in systems with engineered losses, such as coupled photonic waveguides.

Finally, at finite temperature, we find that the presence or absence of state transfer is a signature of the
$\mathcal{PT}$-symmetric or $\mathcal{PT}$-symmetry broken phases, although the dynamics are described by the full, finite-temperature Langevin equation. These results are accessible in a single device through control of the driving strength. 

%


\end{document}